\documentclass{jfm}

\usepackage{graphicx}
\usepackage{newtxtext}
\usepackage{newtxmath}
\usepackage{natbib}
\usepackage{hyperref}
\usepackage{color}
\hypersetup{
    colorlinks = true,
    urlcolor   = blue,
    citecolor  = black,
}

\newcommand{\RomanNumeralCaps}[1]

\newcommand{\Uff}{U_\mathrm{ff}}
\newcommand{\bu}{\boldsymbol{u}}
\newcommand{\bz}{\boldsymbol{z}}
\newcommand{\bomega}{\boldsymbol{\omega}}
\newcommand{\unit}[1]{\,\mathrm{#1}}
\newcommand{\degC}{^\circ\mathrm{C}}
\newcommand{\sinc}{\mathrm{sinc}}

\title{Reynolds number scaling and energy spectra in geostrophic convection}
\author{Matteo Madonia\aff{1}, Andr\'es J. Aguirre Guzm\'an\aff{1}, Herman J.H. Clercx\aff{1} and Rudie P.J. Kunnen\aff{1}\corresp{\email{r.p.j.kunnen@tue.nl}}}
  
\affiliation{\aff{1}Fluids and Flows group, Department of Applied Physics and J. M. Burgers Centre for Fluid Dynamics, Eindhoven University of Technology, P.O. Box 513, 5600 MB Eindhoven, The Netherlands}

\begin{document}
\maketitle

\begin{abstract}
We report flow measurements in rotating Rayleigh--B\'enard convection in the rotationally-constrained geostrophic regime. We apply stereoscopic particle image velocimetry to measure the three components of velocity in a horizontal cross-section of a water-filled cylindrical convection vessel. At a constant, small Ekman number $Ek=5\times 10^{-8}$ we vary the Rayleigh number $Ra$ between $10^{11}$ and $4\times 10^{12}$ to cover various subregimes observed in geostrophic convection. We also include one nonrotating experiment. The scaling of the velocity fluctuations (expressed as the Reynolds number $Re$) is compared to theoretical relations expressing balances of viscous--Archimedean--Coriolis (VAC) and Coriolis--inertial--Archimedean (CIA) forces. Based on our results we cannot decide which balance is most applicable here; both scaling relations match equally well. A comparison of the current data with several other literature datasets indicates a convergence towards diffusion-free scaling of velocity as $Ek$ decreases. However, the use of confined domains leads at lower $Ra$ to prominent convection in the wall mode near the sidewall. Kinetic energy spectra point at an overall flow organisation into a quadrupolar vortex filling the cross-section. This quadrupolar vortex is a quasi-two-dimensional feature as it only manifests in energy spectra based on the horizontal velocity components. At larger $Ra$ the spectra reveal the development of a scaling range with exponent close to $-5/3$, the classical exponent for inertial-range scaling in three-dimensional turbulence. The steeper $Re(Ra)$ scaling at low $Ek$ and development of a scaling range in the energy spectra are distinct indicators that a fully developed, diffusion-free turbulent flow state is approached, sketching clear perspectives for further investigation.
\end{abstract}

\section{Introduction\label{ch:intro}}
Buoyant convection and rotational influence through Coriolis acceleration are two principal features of fluid flows encountered in geophysics and astrophysics. A popular model system to study buoyant convective flows is Rayleigh--B\'enard convection (RBC): a horizontal layer of fluid sandwiched between two parallel plates where the bottom plate is at a higher temperature than the top plate. With the inclusion of background rotation about a vertical axis, the problem of rotating Rayleigh--B\'enard convection (RRBC) encompasses the principal actors, convection and rotation, in a simple, mathematically well-defined system amenable to experimental, numerical and analytical investigation. The interplay of convection and rotation is often studied in the RRBC context \citep{scl13,k21,es23}, which in itself has so far presented a wide variety of possible flow states without invoking additional complexities that may be encountered in geo- and astrophysical flows such as magnetic fields, strong non-Oberbeck--Boussinesq effects, spherical geometries, and more \citep[e.g.,][]{j11,g14,accjknss15,ss20}.

A property shared by the majority of geo- and astrophysical flows is that they are characterised by large Reynolds numbers (ratio of inertial to viscous forces) and at the same time small Rossby numbers (ratio of inertial to Coriolis forces). We thus expect turbulent convective flow with a strong rotational constraint. In recent RRBC investigations it was found that rotation-dominated convection --- named the geostrophic regime after the dominant balance between Coriolis force and pressure gradient --- displays an interesting subdivision into various realisable flow phenomenologies, each with specific scalings of descriptive statistical quantities like the efficiency of convective heat transfer (e.g., \citet{sjkw06,jrgk12,sljvcrka14,en14,csrgka15,cmak20,ldsxz21,mack21}), which is recently reviewed in \citet{k21} and \citet{es23}. Starting from onset of convection, with increasing strength of buoyant forcing one expects to encounter cellular convection, convective Taylor columns (CTCs), plumes and geostrophic turbulence (GT), showing step by step a decreasing rotational constraint \citep{sjkw06,jrgk12,k21,amcock22}. From our prior heat transfer and temperature measurements \citep{cmak20} we could identify an additional transitional state, rotation-influenced turbulence (RIT), that requires further description.

Getting to the geostrophic regime requires dedicated tools: large-scale experimental setups \citep[e.g.,][]{en14,csrgka15,cajk18,cmak20,h20,ldsxz21}, direct numerical simulations (DNSs) on fine meshes \citep[e.g.,][]{sljvcrka14,amcock20} or asymptotically reduced modelling \citep[e.g.,][]{sjkw06,jrgk12,mkjc21}. A complicating factor for experiments in particular is that at the confining sidewall a prominent mode of convection is formed, named the wall mode, boundary zonal flow or sidewall circulation \citep{zghwzaewbs20,zes21,wamcck20,fk20,ldsxz21,ezs22,wmfw22}. This convection mode is the first to become unstable to buoyant forcing, before bulk convection \citep[e.g.,][]{zl09}, and was found to unexpectedly persist deep into the turbulent regimes \citep{mack21}. This complicates the interpretation of results on global flow properties (like the efficiency of convective heat transfer) and their comparison to studies without lateral confinement (i.e. simulations on domains with periodic boundary conditions).

The difficulty of entering the geostrophic regime, particularly for experiments and DNS, leaves it scarcely studied with many open questions remaining, particularly considering the statistical description of this turbulent flow from velocity measurements. Here, we contribute flow measurements from our dedicated experimental setup TROCONVEX \citep{cajk18,cmak20,mack21}. At a constant rotation rate, we choose different strengths of buoyant forcing to scan the subranges of the geostrophic regime. We quantify the strength of turbulence in terms of the Reynolds number and compute kinetic energy spectra to identify the distribution of energy over the spatial scales.

The remainder of this paper is organised as follows. In \S\ref{ch:theory} we introduce the dimensionless input and output parameters of RRBC and list proposed scaling arguments for the Reynolds number. Then, the experimental methods are explained in \S\ref{ch:expmeth}. Results on the measured Reynolds numbers are given in \S\ref{ch:reynolds}, along with a test of the scaling arguments and a comparison to literature data. In \S\ref{ch:spectra} we plot and discuss the computed kinetic energy spectra. We present our conclusions in \S\ref{ch:concl}.

\section{\label{ch:theory}Theoretical background}
We will first define the dimensionless input and output parameters of RRBC in section \ref{ch:params}. Then, the effects of rotation on the stability of convective flow are treated in section \ref{ch:stability}. In section \ref{ch:theoreynolds} we discuss scaling arguments for the Reynolds number proposed in the literature.

\subsection{\label{ch:params}Input and output parameters}
Three dimensionless input parameters are required to describe the flow state. The Rayleigh number $Ra$ expresses the ratio of thermal forcing to dissipation, the Ekman number $Ek$ is the ratio of viscous to Coriolis forces, and the Prandtl number $Pr$ describes the diffusive properties of the fluid. These quantities are defined as
\begin{equation}
Ra=\dfrac{g\alpha\Delta T H^3}{\nu\kappa}\, , \quad Ek=\dfrac{\nu}{2\Omega H^2} \, \quad Pr=\dfrac{\nu}{\kappa} \, ,
\label{eq:parameters}
\end{equation}
where $g$ is gravitational acceleration, $\Delta T$ the applied temperature difference between the plates and $H$ their vertical separation, $\Omega$ is the rotation rate, and $\alpha$, $\nu$ and $\kappa$ are the thermal expansion coefficient, kinematic viscosity and thermal diffusivity of the fluid, respectively. A related parameter that has been used frequently to give an \emph{a priori} indication of the relative importance of buoyant forcing to Coriolis forces \citep[see, e.g.,][]{scl13,ahj20,k21,es23} is the so-called convective Rossby number
\begin{equation}
Ro_c=\dfrac{\Uff}{2\Omega H}=Ek\sqrt{\dfrac{Ra}{Pr}} \, ,
\end{equation}
where the free-fall velocity $\Uff=\sqrt{g\alpha\Delta TH}$ indicates the maximum flow speed that could develop in this convection system.

Two prominent output parameters are the Nusselt number $Nu$ and the Reynolds number $Re$ which indicate the efficiency of convective heat transfer and momentum transfer, respectively. The Nusselt number is defined as
\begin{equation}
Nu=\dfrac{q}{q_\mathrm{cond}}=\dfrac{qH}{\lambda\Delta T}\, .
\end{equation}
The total heat flux $q$ (convection and conduction) is normalized by the conductive flux $q_\mathrm{cond}=\lambda\Delta T/H$ with $\lambda$ the thermal conductivity of the fluid. In absence of convection (a quiescent fluid), $q=q_\mathrm{cond}$ and $Nu=1$; when convection sets in $q>q_\mathrm{cond}$ and $Nu>1$. In our current configuration, with gravity pointing downward, the total heat flux can be expressed as
\begin{equation}
q=\frac{\lambda}{\kappa}\left< w T \right> - \lambda\left<\dfrac{\partial T}{\partial z}\right>\, ,
\label{eq:heatflux}
\end{equation}
where $z$ is the vertical coordinate pointing upward counter to gravity and $w$ is the vertical component of velocity. Angular brackets $\left< \ldots \right>$ denote a suitable average, either a cross-sectional plane average at a certain height $z$ or a volume average over the flow domain. The Reynolds number
\begin{equation}
Re=\dfrac{UH}{\nu}
\end{equation}
introduces a characteristic velocity scale $U$, usually taken to be the root-mean-squared velocity. The ultimate goal is to understand how $Nu$ and $Re$ depend on the input parameters $Ra$, $Pr$, and $Ek$. Power-law scaling relations are expected, e.g., \citet{jkrv12,ksb13,mkjc21}. For nonrotating  convection, the phenomenological theory developed by \citet{gl00,gl01,gl02,gl04} \citep[see also][]{agl09,spgl13} has been very successful in explaining the $Nu(Ra,Pr)$ and $Re(Ra,Pr)$ relations.

\subsection{\label{ch:stability}Stabilisation of convection by rotation}
Rotation tends to stabilise convection. For $Pr>0.68$, linear stability theory \citep{c61,nb65} predicts the critical Rayleigh number $Ra_C$ for onset of convection and the characteristic horizontal size $\ell_C$ of convection near onset (half the most unstable wavelength) as
\begin{equation}
Ra_C=8.70Ek^{-4/3}\, , \quad \ell_C/H=2.41Ek^{1/3} \, .
\label{eq:critical}
\end{equation}
(The onset of convection for $Pr<0.68$ behaves differently \citep{c61,abghv18}; it is out of scope for the current investigation.) Due to this stabilising effect, it is insightful to also consider the degree of supercriticality $Ra/Ra_C$ to get a first impression of the intensity of convection.

\subsection{\label{ch:theoreynolds}Reynolds number scaling in rotating convection}
Several scaling relations for the Reynolds number have been proposed for RRBC flow. They are based on balancing terms in the govering Navier--Stokes equation for an incompressible Oberbeck--Boussinesq fluid, which is \citep[e.g.,][]{c61}
\begin{equation}
\dfrac{\partial\bu}{\partial t}+(\bu\bcdot\bnabla)\bu=-\bnabla p -2\Omega\hat{\bz}\times\bu + \nu\nabla^2 \bu + g\alpha T \hat{\bz} \, ,\quad \bnabla\bcdot\bu=0 \, . \label{eq:ns}
\end{equation}
This equation describes the evolution of velocity $\bu$ as a function of time $t$, where $p$ is the reduced pressure \citep[e.g.,][]{g68,kc02} and $T$ is temperature; $\hat{\bz}$ is the vertical unit vector pointing upward, counter to gravity. Centrifugal buoyancy is excluded here \citep{ha18,ha19}. The equation for vorticity $\bomega=\bnabla\times\bu$ is also invoked in these scaling arguments:
\begin{equation}
\dfrac{\partial\bomega}{\partial t}+(\bu\bcdot\bnabla)\bomega=(\bomega\bcdot\bnabla)\bu+2\Omega\dfrac{\partial\bu}{\partial z}+\nu\nabla^2\bomega - g\alpha\hat{\bz}\times \bnabla T \, .
\end{equation}

One proposed scaling relation balances the viscous, buoyancy and Coriolis terms into the so-called visco--Archimedean--Coriolis (VAC) balance \citep{abncm01,gj06,ksb13}. First, we recall the exact global balance of energy (viscous dissipation equals buoyant production) that can be found by averaging the energy equation $\bu\bcdot\,$(\ref{eq:ns})  over the flow domain \citep{ss90,gl00,agl09}:
\begin{equation}
\nu\left<(\bnabla\bu)^2\right>_V=g\alpha\left<wT\right>_V=\dfrac{\nu^3}{H^4}\dfrac{Ra(Nu-1)}{Pr^2} \, ,
\label{eq:exactbalance}
\end{equation}
where $\left<\ldots\right>_V$ denotes a volume average. For modest convection, geostrophy and the Taylor--Proudman theorem are considered to be broken by viscosity acting on small horizontal scales. The balance of viscous and Coriolis terms in the vorticity equation is scaled by recognising that vertical derivatives $\partial/\partial z$ scale as $1/H$ and horizontal derivatives $\partial/\partial x$ and $\partial/\partial y$ as $1/\ell$, where $\ell\ll H$ is the horizontal scale of convection:
\begin{equation}
2\Omega\dfrac{\partial\bu}{\partial z}\sim \nu \nabla^2\bomega \quad \rightarrow \quad 2\Omega\dfrac{U}{H}\sim\nu\dfrac{U}{\ell^3} \quad \rightarrow \quad \dfrac{\ell}{H}\sim\left(\dfrac{\nu}{2\Omega H^2}\right)^{1/3}=Ek^{1/3} \, .
\label{eq:ellscaling}
\end{equation}
The scaling of $\ell$ is in agreement with onset lengthscale $\ell_C$ of eq. (\ref{eq:critical}).  Using $\bnabla\sim 1/\ell$ (i.e. dominated by the horizontal derivatives) we can scale the first term in eq. (\ref{eq:exactbalance}) as $\nu U^2/\ell^2$, then, using eq. (\ref{eq:ellscaling}), rewrite it in terms of the Reynolds number as
\begin{equation}
\nu\dfrac{U^2}{\ell^2}\sim\dfrac{\nu^3}{H^4}\dfrac{Ra(Nu-1)}{Pr^2} \quad \rightarrow \quad Re_\mathrm{VAC}\sim \dfrac{Ra^{1/2}Ek^{1/3}(Nu-1)^{1/2}}{Pr} \, .
\label{eq:vacscaling}
\end{equation}

A second proposed scaling is built on a balance of Coriolis--inertial--Archimedean (CIA) forces \citep{ip82,abncm01,j11,gcs19,ahj20}. For a fully turbulent flow, effects of viscosity are negligible and Coriolis and inertial forces are balanced in the vorticity equation:
\begin{equation}
2\Omega\dfrac{\partial\bu}{\partial z}\sim (\bu\bcdot\bnabla)\bomega \quad \rightarrow \quad 2\Omega\dfrac{U}{H}\sim\dfrac{U^2}{\ell^2} \quad \rightarrow \quad \dfrac{\ell}{H}\sim\left(\dfrac{U}{2\Omega H}\right)^{1/2}=Ro^{1/2} \, ,
\label{eq:turblength}
\end{equation}
where the horizontal length scale $\ell$ now scales as the square root of the Rossby number $Ro=U/(2\Omega H)$ based on the velocity scale $U$ \citep{gcs19}. Then, inertia and buoyancy are balanced in the energy equation:
\begin{equation}
\bu\bcdot((\bu\bcdot\bnabla)\bu) \sim g\alpha\bu\bcdot(T\hat{\bz}) \quad \rightarrow \quad \dfrac{U^3}{\ell}\sim \dfrac{\nu^3}{H^4}\dfrac{Ra(Nu-1)}{Pr^2} \, ,
\label{eq:ucube}
\end{equation}
where (\ref{eq:exactbalance}) has been used invoking a volume average. The final scaling relation, combining eqs. (\ref{eq:turblength}) and (\ref{eq:ucube}), can be rewritten in terms of $Re$ as
\begin{equation}
Re_\mathrm{CIA}\sim\dfrac{Ra^{2/5}Ek^{1/5}(Nu-1)^{2/5}}{Pr^{4/5}}\, .
\label{eq:ciascaling}
\end{equation}

What remains to be determined is the dependence of the Nusselt number $Nu$ on the input parameters. For rotating convection this relation is far from complete, with a variety of subregimes opening up in the geostrophic regime based on recent numerical and experimental evidence \citep{sjkw06,jrgk12,en14,csrgka15,cajk18,cmak20,ldsxz21,k21}. Two theoretical scalings have been proposed. The first \citep{ksnha09,ksa12} is a rotating equivalent of Malkus's marginal-stability argument for the thermal boundary layer \citep{m54}: the Rayleigh number $Ra_\delta$ based on the thickness $\delta$ of the thermal boundary layer exceeds its critical value $Ra_{\delta,C}\sim Ek_\delta^{-4/3}$ with an Ekman number also based on $\delta$. With the assumption of a temperature drop $\Delta T_\delta\sim\Delta T$ over a thermal boundary layer that is purely conductive ($q\sim k\Delta T/\delta$), \citet{ksnha09,ksa12} find
\begin{equation}
Nu\sim\left(\dfrac{Ra}{Ra_C}\right)^3\sim Ra^3 Ek^4 \, .
\end{equation}
The second scaling relation is a rotating equivalent of Spiegel's argument that under vigorously turbulent conditions the heat flux $q$ should become independent of the diffusive fluid properties $\nu$ and $\kappa$ \citep{s71}. \citet{s79} and \citet{jkrv12} show that, if rotation is included, the only combination of parameters leading to diffusion-free total heat flux $q$ is
\begin{equation}
Nu-1\sim \dfrac{Ra^{3/2}Ek^2}{Pr^{1/2}} \, .
\label{eq:julienscaling}
\end{equation}
Numerical and experimental evidence points towards the asymptotic validity of the latter scaling (\ref{eq:julienscaling}) \citep{jkrv12,sljvcrka14,bmjag21}, although the presence of no-slip walls \citep{sljvcrka14,kopvl16,pjms17,amcock21} and the significant heat flux contribution of the wall mode near sidewalls in confined convection \citep{wamcck20,zghwzaewbs20,zes21,fk20,ldsxz21,ezs22} preclude observation of the pure scaling law.

Upon insertion of (\ref{eq:julienscaling}) into (\ref{eq:ciascaling}) we obtain in dimensionless and dimensional form that
\begin{equation}
Re\sim \dfrac{Ra Ek}{Pr} \quad \rightarrow \quad U\sim\dfrac{g\alpha\Delta T}{2\Omega} \, ,
\label{eq:diffusionfreescaling}
\end{equation}
i.e., the velocity scale $U$ is now also diffusion-free \citep{ahj20}.

\section{\label{ch:expmeth}Experimental methods}
The experimental setup used for this study is TROCONVEX, a large-scale rotating convection apparatus designed for the study of the geostrophic regime of convection. The design considerations with regard to the accessible parameter range for a given setup have been discussed in detail in \citet{cajk18}. Details on the setup for heat transfer measurements are given in \citet{cmak20}. Here we perform optical flow measurements using stereoscopic particle image velocimetry \cite[stereo-PIV;][]{p00spiv,rwwk07}. The arrangement has been explained in \citet{mack21}; here we repeat the most important parts.

The convection cell is an upright cylinder of height $H=2\unit{m}$ and radius $R=0.195\unit{m}$ filled with water. Its diameter-to-height aspect ratio is $\Gamma=2R/H=0.195$. The bottom plate is made of copper and is electrically heated to be at a controlled temperature $T_b$. Likewise, the temperature of the copper top plate, equipped with a double spiral groove for cooling liquid circulation, is controlled to a temperature $T_t$ by a combination of a chiller and a thermostated bath. The mean temperature $T_m=(T_b+T_t)/2$ is kept at a constant $31\degC$, so that $Pr=5.2$. The convection cell is placed on a rotating table. We apply constant rotation $\Omega=1.93\unit{rad/s}$, corresponding to $Ek=5.00\times 10^{-8}$. The temperature difference $\Delta T=T_b-T_t$ is changed between experiments to vary $Ra$. Additionally, one nonrotating experiment is included for reference. The operating conditions for the various experiments are summarised in table \ref{ta:exps}.

\begin{table}
\begin{center}
\def~{\hphantom{0}}
\begin{tabular}{cccccc}
$\Delta T$ ($\degC$)  & $Ra/10^{12}$ & $Ek\times 10^{8}$ & $Ra/Ra_C$   & $Re_u/10^3$     & $Re_w/10^3$     \\[3pt]
$~0.50$               & $0.108$      & $5.00$            & $~2.29$     & $~3.10\pm 0.21$ & $~2.31\pm 0.27$ \\
$~1.00$               & $0.216$      & $5.00$            & $~4.57$     & $~3.09\pm 0.25$ & $~2.18\pm 0.20$ \\
$~2.00$               & $0.432$      & $5.00$            & $~9.14$     & $~3.88\pm 0.20$ & $~2.63\pm 0.31$ \\
$~3.00$               & $0.648$      & $5.00$            & $13.7~$     & $~5.80\pm 0.80$ & $~3.43\pm 0.31$ \\
$~5.00$               & $1.08~$      & $5.00$            & $22.9~$     & $~7.93\pm 0.95$ & $~5.60\pm 0.48$ \\
$10.0~$               & $2.16~$      & $5.00$            & $45.7~$     & $12.88\pm 0.88$ & $~8.66\pm 0.46$ \\ 
$20.0~$               & $4.32~$      & $5.00$            & $91.4~$     & $17.67\pm 0.92$ & $12.90\pm 1.16$ \\[3pt] 
$~3.00$               & $0.648$      & $\infty$          & $O(10^{8})$ & $10.34\pm 1.61$ & $13.66\pm 2.59$ \\ 
\end{tabular}
\caption{\label{ta:exps}Parameter values for the experiments. In all cases $\Gamma=0.195$ and $Pr=5.2$. Output parameters are the Reynolds numbers $Re_u$ and $Re_w$ based on horizontal and vertical root-mean-square velocities, respectively. For the nonrotating experiment (bottom row) the parameter $Ra/Ra_C$ is not explicitly stated. It is large given that without rotation $Ra_C$ is only $O(10^3)$; quantitative comparison to the rotating cases is not meaningful.}
\end{center}
\end{table} 

Optical access is facilitated by a custom-made water-filled prism mounted around the cylinder, significantly reducing refraction on the cylinder wall. It allows for horizontal crossing of a laser light sheet with a thickness of about $3.5\unit{mm}$ at midheight $z/H=0.5$. The laser pulses at frequencies of $7.5$ or $15\unit{Hz}$ depending on the typical flow speeds. The water is seeded with polyamid particles with a nominal diameter of $5\unit{\umu m}$. Two $5\unit{MPixel}$ cameras (Jai SP-500M-CXP2) equipped with Scheimpflug adapters are each placed at a stereoscopic angle of approximately $45^\circ$ with the vertical, so that stereo-PIV can be applied \citep{p00spiv,rwwk07}. We can measure the full velocity vectors $\bu=(u,v,w)$ in the laser light sheet plane, resulting in a grid of vectors with a separation of $3.2\unit{mm}$ in both horizontal directions that fits $122$ vectors in the cylinder diameter. Between $3000$ and $9000$ vector fields are evaluated per experiment for a measurement duration ranging from $200$ to $600\unit{s}$.

Results are presented in terms of Reynolds numbers. Horizontal and vertical velocities are treated separately. The Reynolds number $Re_u$ is based on the characteristic horizontal velocity $U_h=\sqrt{\left<u^2\right>+\left<v^2\right>}$. Here $\left<\ldots\right>$ either denotes averaging over time and over circular shells to obtain radial profiles, or averaging over time and over the full cross-sectional area excluding the wall mode near the sidewall (to be more precisely defined later on) to get to a single representative value per experiment. In this azimuthal averaging, we divide the section in 65 concentric circular shells. Each shell represents a ring of thickness $3\unit{mm}$ comparable to the resolution of our PIV vector field spacing ($3.2\unit{mm}$). Similarly, $Re_w$ is based on the characteristic velocity scale $W=\sqrt{\left<w^2\right>}$. In all cases $\left<u\right>\approx\left<v\right>\approx\left<w\right>\approx 0$. These Reynolds numbers can also be interpreted as the root mean square (rms) velocities normalised with the viscous velocity scale $U_\nu=\nu/H$. Likewise, we consider the rms value of the vertical component of vorticity, $\omega_z^\mathrm{rms}$.

An important consideration that is encountered in rotating convection experiments is the effect of centrifugal buoyancy \citep{ha18,ha19}. We use the Froude number $Fr=\Omega^2R/g$ to quantify the ratio of maximal centrifugal to gravitational acceleration. Since centrifugal acceleration is negligibly small in most geophysical and astrophysical applications, its effects \citep[as studied in detail by][]{ha18,ha19} should preferably be kept to a minimum in experiments, i.e. $Fr\ll 1$. Here, $Fr=0.074$ for all rotating experiments. Such $Fr$ did not lead to significant up/down asymmetry in our sidewall temperature measurements \citep{cmak20}; we expect that centrifugal buoyancy is negligibly small in these experiments, too.

\begin{figure}
\includegraphics[width=\textwidth]{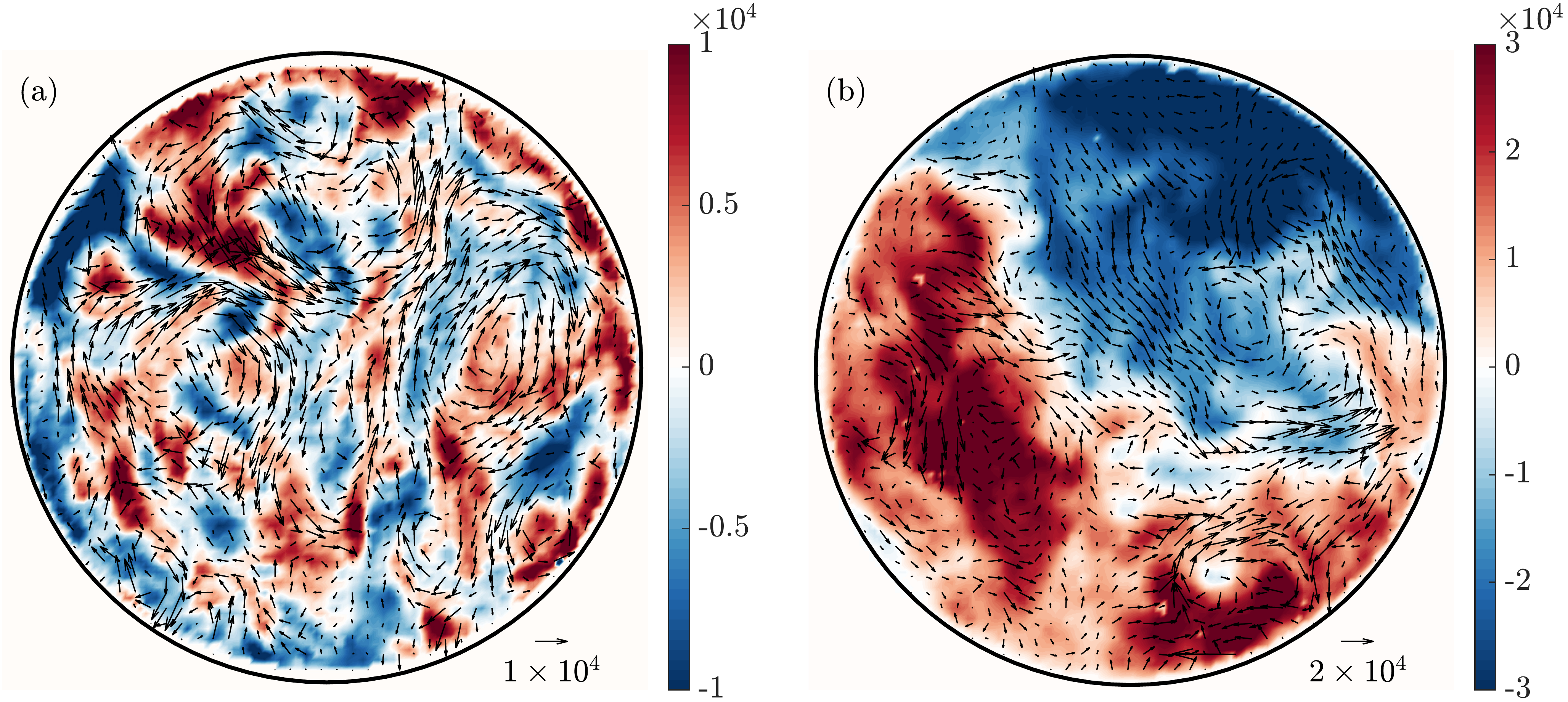}
\caption{\label{fi:snaps}Instantaneous velocity snapshots at $Ra=6.48\times 10^{11}$: (a) with rotation, $Ek=5.00\times 10^{-8}$, (b) no rotation, $Ek=\infty$. The arrows display the horizontal velocity components; for clarity only one ninth of the total number of vectors is displayed. The background colour indicates the vertical velocity component. Velocities are normalised with the viscous velocity $\nu/H=3.87\times 10^{-7}\unit{m/s}$.}
\end{figure}

To give a first impression of the measurement results, we plot two instantaneous velocity field snapshots at $Ra=6.48\times 10^{11}$, with and without rotation, in figure \ref{fi:snaps}. Velocities are normalised with the viscous velocity scale $\nu/H$, which is $3.87\times 10^{-7}\unit{m/s}$ at this mean temperature. As such, the magnitude of the normalised velocity components in these plots can be interpreted as displaying the local instantaneous Reynolds number. It is clear that rotation, present in panel (a) but absent in panel (b), has a large effect on the overall flow structuring. Under rotation the flow tends to organise into structures considerably smaller than the diameter of the cylinder, while without rotation the global organisation is as large as the cross-section. We have considered the characteristic size of the flow features before in \citet{mack21}, where we could see that the correlation length for vertical velocity increases as $Ra$ grows. In the rotating experiments we can also observe intense vertical motion near the sidewall due to the convective wall mode \citep{wamcck20,zghwzaewbs20,zes21,fk20,ldsxz21,ezs22}. For roughly half of the circumference we find upward flow close to the sidewall; the other half downward. In the nonrotating experiment (figure \ref{fi:snaps}(b)) the large-scale circulation is observed \citep[e.g.,][]{agl09}: a vertical convection roll filling the domain with one half upward and one half downward flow.

\section{\label{ch:reynolds}Reynolds number results}

\subsection{Radial profiles}
\begin{figure}
\includegraphics[width=\textwidth]{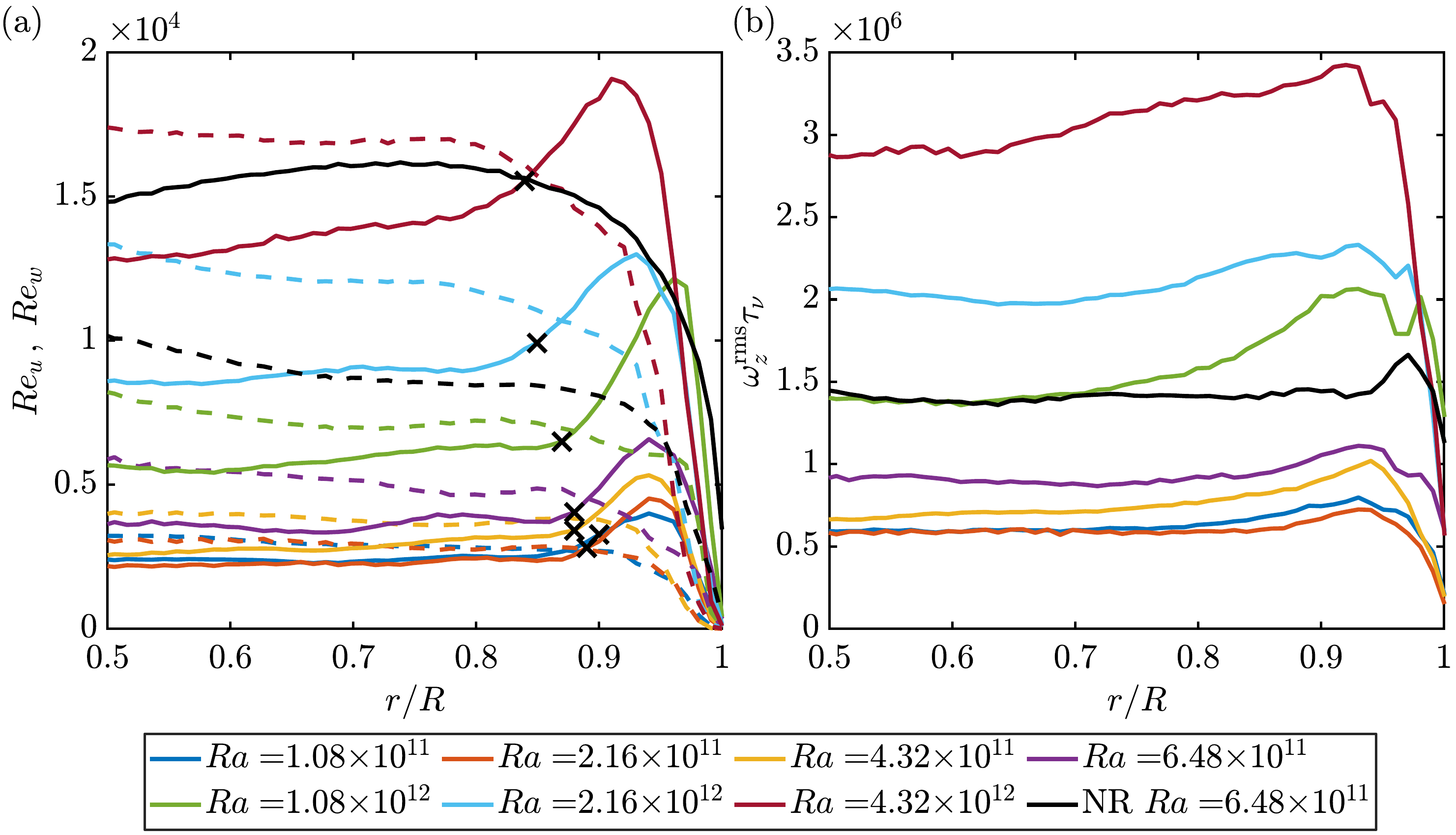}
\caption{\label{fi:avg_vel_vort}(a) Radial dependence of $Re_u$ (dashed lines) and $Re_w$ (solid lines) for different $Ra$. Black crosses indicate the beginning of the sidewall boundary layer following \ref{eq:wall_deWit}. (b) Radial dependence of $\omega_z^\mathrm{rms}$ for different $Ra$, normalised using the viscous time scale $\tau_{\nu}=H^2/\nu$. The legend entry `NR' refers to the nonrotating experiment. All the quantities displayed in this figure are only shown from half the cylinder radius onward for clarity.}
\end{figure}
We first consider radial profiles of the Reynolds numbers $Re_u$ and $Re_w$, plotted in figure \ref{fi:avg_vel_vort}(a). Both quantities show an overall trend of larger values for larger $Ra$, as expected, with the exception of the two lowest $Ra$ cases that are almost identical. The distinction between an inner and a near-wall outer portion of the domain is clear from both quantities: both profiles suggest a generally constant value that is largely independent of the radial position for the bulk part, away from the lateral sidewall, and a change of behavior close to this sidewall. There we see an increase in $Re_w$, a clear indication of the presence of vigorous convection near the sidewall in the wall mode. The black crosses in figure \ref{fi:avg_vel_vort}(a) indicate the approximate thickness of the sidewall boundary layer following the empirical relation
\begin{equation}
\label{eq:wall_deWit}
\delta_{w_\mathrm{rms}}/R=a Ra^{0.15\pm 0.02} \, ,
\end{equation}
based on capturing the radial extent of the near-wall $w_\mathrm{rms}$ peak; $a=(3\pm 1)\times 10^{-3}$ is reported by \citet{wamcck20}. Here we use $a=2\times 10^{-3}$ as the prefactor and $0.15$ as the exponent for $Ra$. This definition agrees well with our data as it indicates the extent of the sidewall layer, in both $Re_w$ and $Re_u$. For the former it is at the base of the near-wall peak, for the latter we see the beginning of the decay to zero.% We shall return to the sidewall layer in section \ref{ch:thickness_of_sidewall_circulation}.

The non-rotating case (black curves in figure \ref{fi:avg_vel_vort}(a)) shows three main points of interest. First, $Re_w$ displays an immediate decay to zero close to the sidewall without any previous increase, contrary to the rotating cases. Second, without rotation we measure significantly higher values of both $Re_u$ and $Re_w$ relative to the rotating case at the same $Ra$ (compare black and purple curves in figure \ref{fi:avg_vel_vort}(a)), a clear indication that the strong influence of rotation makes the flow less turbulent, suppressing both vertical and horizontal rms velocities. Third, the rms vertical velocity is now considerably more pronounced than the horizontal component, unlike the rotating cases. We will discuss flow anisotropy later in this paper.

In figure \ref{fi:avg_vel_vort}(b) we plot $\omega_z^\mathrm{rms}$ normalised using the viscous time scale $\tau_{\nu}=H^2/\nu$. The same trends as for $Re_w$ are reflected here, with the two cases with lowest $Ra$ being almost identical and profiles that increase for higher $Ra$. Also here we see higher values for the non-rotating case compared to its corresponding rotating counterpart, even though the difference is less than the one we see for vertical velocities. The non-rotating case also shows a localised peak close to the sidewall, while the rotating cases have a wider radial region where the vorticity increases before dropping down at the wall.

All the quantities displayed in figure \ref{fi:avg_vel_vort} are shown from half the cylinder radius onward for clarity. The inner part keeps showing a constant behavior down to approximately $\frac{1}{10}R$, a circle of around $20\unit{mm}$ around the axis of the cylinder where the scarcity of velocity vectors prevents meaningful radial binning and azimuthal averaging.

\begin{figure}
\includegraphics[width=\textwidth]{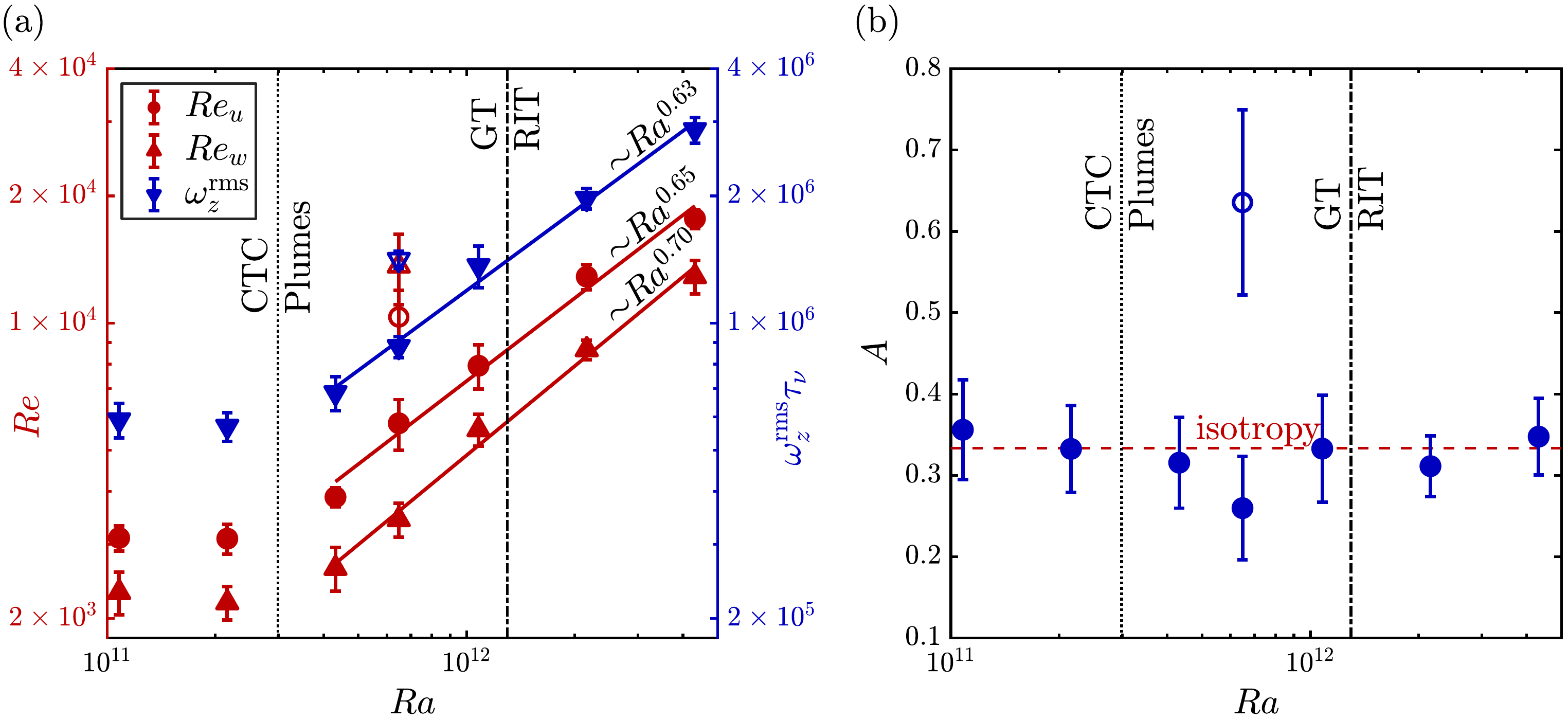}
\caption{\label{fi:avg_vel_vor}(a) Dependence on $Ra$ of Reynolds numbers based on horizontal ($Re_u$) and vertical ($Re_w$) velocity components (red, left ordinate) and vertical vorticities ($\omega_z^\mathrm{rms}$; blue, right ordinate). Vorticity is normalised with the viscous timescale $\tau_\nu=H^2/\nu$. The solid lines represent power-law fits $Re_u\sim Ra ^{0.65 \pm 0.07}$, $Re_w\sim Ra ^{0.70 \pm 0.06}$ and $\omega_z^\mathrm{rms} \sim Ra^{0.63 \pm 0.05}$. (b) Kinetic energy anisotropy $A=W^2 / (U_h^2+W^2)$ versus $Ra$. For isotropy $A=\tfrac{1}{3}$ (red dashed line). In both panels the dotted and the dash-dotted lines represent the transitional $Ra$ between CTCs and plumes and between GT and RIT, respectively, while open symbols represent the values for the nonrotating case.}
\end{figure}

The quantities $Re_u$, $Re_w$ and $\omega_z^\mathrm{rms}$ are approximately constant in the inner part of the cylinder. From the profiles in figure \ref{fi:avg_vel_vort} we can extract for every $Ra$ a mean value of that quantity in the bulk. This bulk average is obtained by excluding the wall zone as defined by equation (\ref{eq:wall_deWit}) as well as excluding a circle of radius $20\unit{mm}$ from the cylinder axis, an area where the azimuthal averaging does not give trustworthy data, as argued above. In figure \ref{fi:avg_vel_vor}(a) we plot these averaged data as a function of $Ra$, with error bars that represent the standard deviations of these means. We also give the corresponding flow states as inferred from our earlier heat transfer measurements \citep{cmak20}.

As mentioned before, the two lower $Ra$ cases, both in the CTC regime, show very similar values. From the plumes regime onward, all the quantities display a steeper trend, that is reasonably constant between the plumes/GT and RIT states: $Re_u\sim Ra ^{0.65 \pm 0.07}$, $Re_w\sim Ra ^{0.70 \pm 0.06}$ and $\omega_z^\mathrm{rms} \sim Ra^{0.63 \pm 0.05}$. These exponents are notably larger than the nonrotating trend $Re\sim Ra^{0.44}$ that has been measured in various experiments and employing different methods, as summarised by \citet{agl09}. They are even larger than the ``ultimate" diffusion-free scaling $Re\sim Ra^{1/2}$ for nonrotating convection \citep{k62,s71,gl02}. Another clear difference between the current nonrotating and rotating experiments is that without rotation the values are higher than with rotation at the same $Ra$.

The vorticity scaling $\omega_z^\mathrm{rms}\sim Ra ^{0.63}$ nicely follows that of horizontal velocity. From this information we can infer that  the characteristic horizontal length scale $\ell$ does not change much with $Ra$, employing the estimate $\omega_z\sim U_h/\ell$. This $Ra$-independent characteristic scale for rapidly rotating convection is one of the starting points of the asymptotically reduced equations \citep[e.g.][]{sjkw06,jrgk12} that describe the flow in the limit $Ek\rightarrow 0$ and that have provided a lot of insights into geostrophic convection. We have measured lengthscales before in \citet{mack21}, verifying the $Ra$-independence of $\ell$.

To study the degree of anisotropy in the bulk flow, we plot in figure \ref{fi:avg_vel_vor}(b) the kinetic energy anisotropy for each case, defined as
\[ A= \dfrac{W^2}{U_h^2+W^2} \, .\]
It describes the fraction of total kinetic energy found in vertical motions. In the isotropic case, where the energy is equally distributed among the three components, this value would be $\frac{1}{3}$. As we see from figure \ref{fi:avg_vel_vor}(b), the non-rotating case is far away from that value, while the rotating cases display values very close to the isotropic one, with the possible exception of the case with $Ra=6.48 \times 10^{11}$, possibly a point of transition between the plumes and geostrophic-turbulence (GT) states of the geostrophic regime, for which no transition relation is currently available. Here, rotation thus suppresses the large anisotropy of nonrotating convection, ending up in near-isotropy.

\subsection{\label{ch:test}Test of force balance scaling relations}
\begin{figure}
\includegraphics[width=\textwidth]{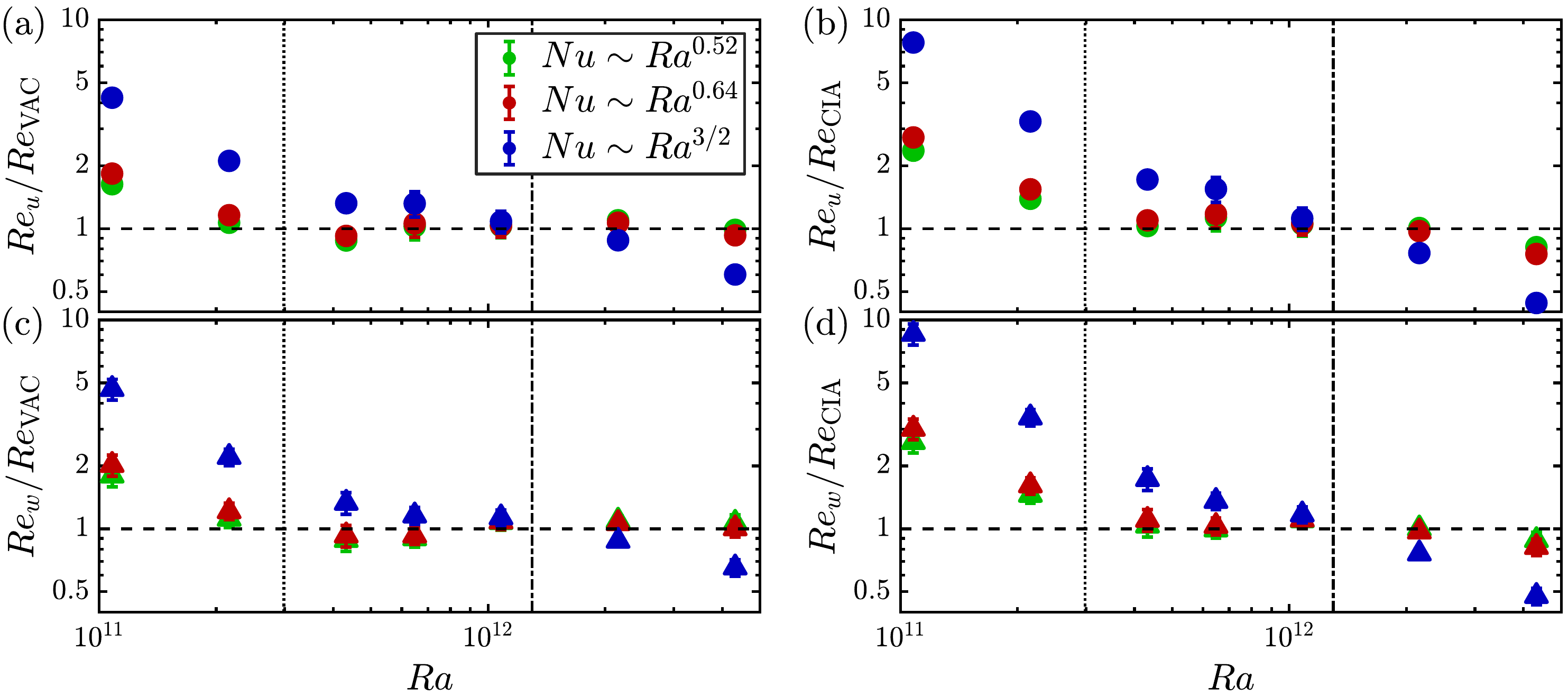}
\caption{\label{fi:testscaling}Test of $Re(Ra)$ scaling with VAC (eq. (\ref{eq:vacscaling})) and CIA (eq. (\ref{eq:ciascaling})) force balance arguments, for (a,b) $Re_u$ and (c,d) $Re_w$. Three different $Nu(Ra)$ relations are invoked: $Nu\sim Ra^{0.52}$ (green; for RIT) and $Nu\sim Ra^{0.64}$ (red; for plumes/GT) based on our heat transfer measurements in the same setup \citep{cmak20}; and $Nu\sim Ra^{3/2}$ (blue) following eq. (\ref{eq:julienscaling}).}
\end{figure}

We can use the measured characteristic velocities to test compliance (in terms of $Ra$-scaling) with the scaling relations (\ref{eq:vacscaling}) and (\ref{eq:ciascaling}), respectively based on the VAC and CIA force balances. We graphically test the compliance in figure \ref{fi:testscaling} by plotting $Re/Re_\mathrm{VAC}$ and $Re/Re_\mathrm{CIA}$. Input on the $Nu(Ra)$ scaling is required. Here, we use our prior heat transfer results obtained in the same experimental setup by \citet{cmak20}, viz. $Nu\sim Ra^{0.64}$ for the plumes/GT range and $Nu\sim Ra^{0.52}$ for the RIT range. It must be emphasised that these measured relations may be affected by the heat transfer contribution of the sidewall circulation \citep{wamcck20,zghwzaewbs20,ldsxz21}. We also include the ultimate scaling $Nu\sim Ra^{3/2}$ as per equation (\ref{eq:julienscaling}). The VAC scaling test is included for this diffusion-free relation, too, despite the incompatibility of combining a diffusion-free $Nu$ scaling with a force balance that explicitly includes the viscous force. The $Re$ scaling relations (\ref{eq:vacscaling}) and (\ref{eq:ciascaling}) do not have numerical prefactors. To make these plots, we have chosen the prefactors such that the points are at the same level on average for ease of comparison.

For both $Re_u$ and $Re_w$, it is observed that either VAC or CIA scaling does a good job of describing the results, provided our earlier experimental $Nu(Ra)$ relations are employed. We see that both relations $Nu\sim Ra^{0.52}$ and $Nu\sim Ra^{0.64}$ lead to comparable results and are equally valid in the plumes/GT and RIT ranges. Based on these plots and on the current data, we cannot decide which scaling (VAC or CIA) is most appropriate. This is in line with the analysis of \citet{h20}, who similarly could not reach a conclusion based on measurements using laser Doppler velocimetry. We can insert these numbers into the scaling relations (\ref{eq:vacscaling}) and (\ref{eq:ciascaling}) for a quantitative comparison. Using $Nu\sim Ra^{0.64}$ for the plumes/GT range, we arrive at the prediction $Re \sim Ra^{0.82}$ from VAC scaling and $Re \sim Ra^{0.66}$ from the CIA balance. When we use $Nu\sim Ra^{0.52}$ as obtained for the RIT range, we find $Re \sim Ra^{0.76}$ for VAC and $Re \sim Ra^{0.61}$ for CIA. Our exponents for the velocity scalings from figure \ref{fi:avg_vel_vor} are in this range, perhaps a bit closer to the CIA trend than to the VAC scaling, but certainly not giving a conclusive answer either. We are probably in a state where both inertial and viscous forces play a role. The flow is turbulent enough that inertial forces are relevant, but not yet turbulent enough to reach the diffusion-free scaling $Re\sim Ra$ of eq. (\ref{eq:diffusionfreescaling}).

\begin{figure}
\includegraphics[width=\textwidth]{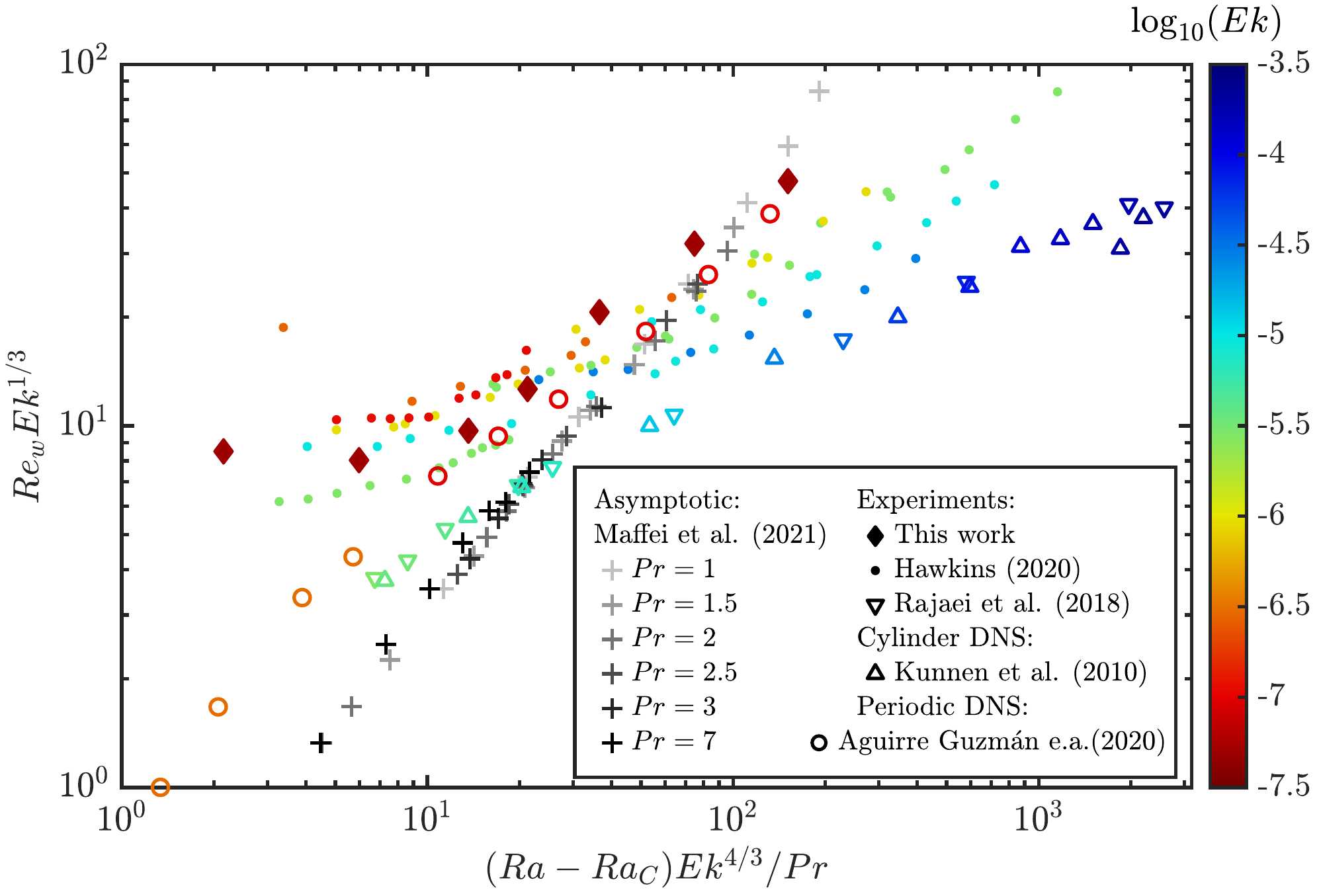}
\caption{\label{fi:Reynolds}Comparison of current results for $Re_w$ with published results. Plusses: data from \citet{mkjc21}, colour coded by $Pr$, results from asymptotically reduced model simulations on a horizontally periodic domain. Other symbols colour coded by $Ek$ (see colour bar). Diamonds: current results. Dots: experiments of \citet{h20}. Up triangles: DNS in a cylinder \citep{kgc10jfm}. Down triangles: experiments of \citet{rakc18}. Circles: DNS in a horizontally periodic domain \citep{amcock20}.}
\end{figure}

To further interpret our data and compare to previously published results, we collect several reported datasets for $Re_w$ in figure \ref{fi:Reynolds}. We plot $\widetilde{Re}_w=Re_w Ek^{1/3}$ as a function of $(\widetilde{Ra}-\widetilde{Ra}_C)/Pr=(Ra-Ra_C)Ek^{4/3}/Pr$, both inspired and necessitated by the results of \citet{mkjc21} from asymptotic model simulations. In their asymptotic equation $Ek\rightarrow 0$ so $Ek$ itself remains undefined. Furthermore, using this plot convention they obtained a good collapse of their data spanning a range of $Pr$ and $\widetilde{Ra}$, see the plus symbols in the plot. Their simulation domain is a Cartesian box with periodic boundary conditions in the horizontal directions and stress-free walls on bottom and top. The other datasets are from DNS and experiments, where the corresponding Ekman number is colour coded. The current results are included with diamonds. The experimental results by \citet{h20} cover a significant range of $Ra\simeq 10^8-2\times 10^{11}$ and $Ek\simeq 10^{-7}-3\times 10^{-5}$; they are the reported RRBC Reynolds number data closest to the current study in terms of $Ra$ and $Ek$ and are included with dots. Numerical and experimental results corresponding to a smaller cylindrical convection cell at $Ra\approx 10^9$ and $Ek\simeq 3\times 10^{-6}-10^{-3}$ taken from \citet{kgc10jfm} and \citet{rakc18} are included with up and down triangles. Finally, the DNS study by \citet{amcock20} on a horizontally periodic domain with no-slip walls on bottom and top at $Ra=5\times 10^9-10^{12}$ and $Ek\approx 10^{-7}$ is included with open circles. From this figure, we can make several observations.

First, the plateauing at low $Ra$ that we observed in figure \ref{fi:avg_vel_vor}(a) is also observed in the data of \citet{h20}. On the contrary, the simulation results on horizontally periodic domains \citep{amcock20,mkjc21} do not show such a plateau. The principal distinguishing property is the presence or absence of a sidewall. We know that the sidewall circulation generates jet-like intrusions from the sidewall region into the bulk \citep{mack21}. It is thus plausible that the observed plateauing of $Re_w$ is caused by the dominance of the jets emerging from the sidewall circulation over the columnar convection in the central region. Note that the difference between the DNS of \citet{amcock20} and the asymptotic model of \citet{mkjc21} at low $\widetilde{Ra}=RaEk^{4/3}$ is related to the applied boundary conditions: in this range, paradoxically, no-slip boundaries lead to higher $Nu$ (and thus also higher $Re_w$) than stress-free boundaries at otherwise identical parameter values \citep{st10,sljvcrka14,kopvl16,pjms17}. This effect has been ascribed to the prevalence of Ekman pumping as a source of vertical transport, present for no-slip plates but absent for stress-free.

Second, the Reynolds number scaling depends sensitively on the magnitude of $Ra$ and $Ek$. The smaller-scale cylinder investigations of \citet{kgc10jfm} and \citet{rakc18} at moderate $Ra\approx 10^9$ and $Ek$ down to $3\times 10^{-6}$ display a significantly shallower slope than the larger cylinders (larger $Ra$ and smaller $Ek$) of \citet{h20} and the current study. There is definitely a different scaling upon entering the geostrophic regime of rotating convection. In fact, these four studies together display a gradual transition as a function of both $Ra$ and $Ek$. A steeper scaling can be seen as $Ek$ is reduced, while concomitantly increasing $Ra$ to retain turbulent convection, save for the added complications due to the sidewall circulation.

Third, the smallest-$Ek$ datapoints (orange and red) display a trend towards convergence to the asymptotic results for $(Ra-Ra_C)Ek^{4/3}/Pr \gtrsim 50$ and $Re_w Ek^{1/3}\gtrsim 20$. Notably, results of completely different origin (experiments, DNS and asympotic models; horizontally periodic domains and confined cylinders) display convergence towards a common scaling behaviour. This can be taken as a strong indicator that the ``ultimate" diffusion-free scaling is approached. A linear scaling in this diagram directly corresponds to the relation expressed in eq. (\ref{eq:diffusionfreescaling}) where the velocity scale becomes independent of both diffusive parameters and domain-specific parameters like its height $H$. The convergence is not yet fully achieved; extra datapoints at even smaller $Ek$ are required to corroborate this, but tough to realise experimentally or numerically.

\section{\label{ch:spectra}Energy spectra}
To compute the energy spectra we employ two-dimensional fast Fourier transforms of the velocity field, after padding the vector fields with zeros up to 512 grid points per dimension to avoid artifacts due to implied periodicity in the Fourier transform \citep{r08}. The two-dimensional data is averaged over circular shells to obtain one representative curve independent of azimuthal angle. Since adjacent velocity vectors are $\Delta x=3.2\unit{mm}$ apart, the smallest wavenumber after zero padding is $k_\mathcal{L}=2\pi/\mathcal{L}=2\pi/(512\Delta x)\approx 3.83\unit{m^{-1}}$, and the maximum relevant $k$ value (Nyquist wavenumber) is $k_{2\Delta x}=2\pi/(2\Delta x)\approx 982 \unit{m^{-1}}$. All the $k$ values are integer multiples of $k_\mathcal{L}$. Nevertheless, the smallest physically relevant $k$ is determined by the diameter $D$ of the cylinder: that would be given by $k_{D}=2\pi /D\approx 16.1 \unit{m^{-1}}$, but since all the wavenumbers are integer multiples of $k_\mathcal{L}$, our first relevant value is $k_\mathrm{min}\approx 19.2 \unit{m^{-1}}$, in dimensionless form $k_\mathrm{min}H\approx 38.4$.

The maximum relevant $k$ requires more consideration. Previous literature \citep{fcs04,s06} has shown that the window averaging applied in PIV measurements acts as a low-pass filter in Fourier space with a cut-off wavenumber that depends on the size of the measurement window $X$ (in our case $X=2\Delta x=6.4\unit{mm}$). The PIV filtering amounts to a multiplication in Fourier space of the real signal with a squared cardinal sine function
\begin{equation}
\sinc^2(kX/2)=\left(\frac{\sin{kX/2}}{kX/2}\right)^2\, .
\end{equation}
We can then correct for this filtering, with corrected energy spectra becoming
\begin{equation}
\label{eq:e_real}
E_\mathrm{corr}=\frac{E_\mathrm{meas}}{\sinc^2(kX/2)}\,,
\end{equation}
where $E_\mathrm{meas}$ is the uncorrected spectrum determined directly from the PIV data and $E_\mathrm{corr}$ is the corrected spectrum. However, given that $\sinc^2(kX/2)\rightarrow 0$ as $kX/2\rightarrow \pi$, direct application of eq. (\ref{eq:e_real}) may cause practical difficulties: measurement noise will be severely amplified when approaching the zero of the denominator. To deal with this, we employ the rule of thumb proposed by \citet{fcs04} to cut off the spectra at a maximum $k_\mathrm{max}=2.8/X$. At this $k$ value the filter function $\sinc^2(k_{max}X/2)\approx 0.5$, i.e. the spectrum is cut off at the wavenumber $k_\mathrm{max}$ where PIV filtering reduces the measured spectral amplitude by $50\%$. In our case $k_\mathrm{max}H= 874$.

\begin{figure}
\includegraphics[width=\textwidth]{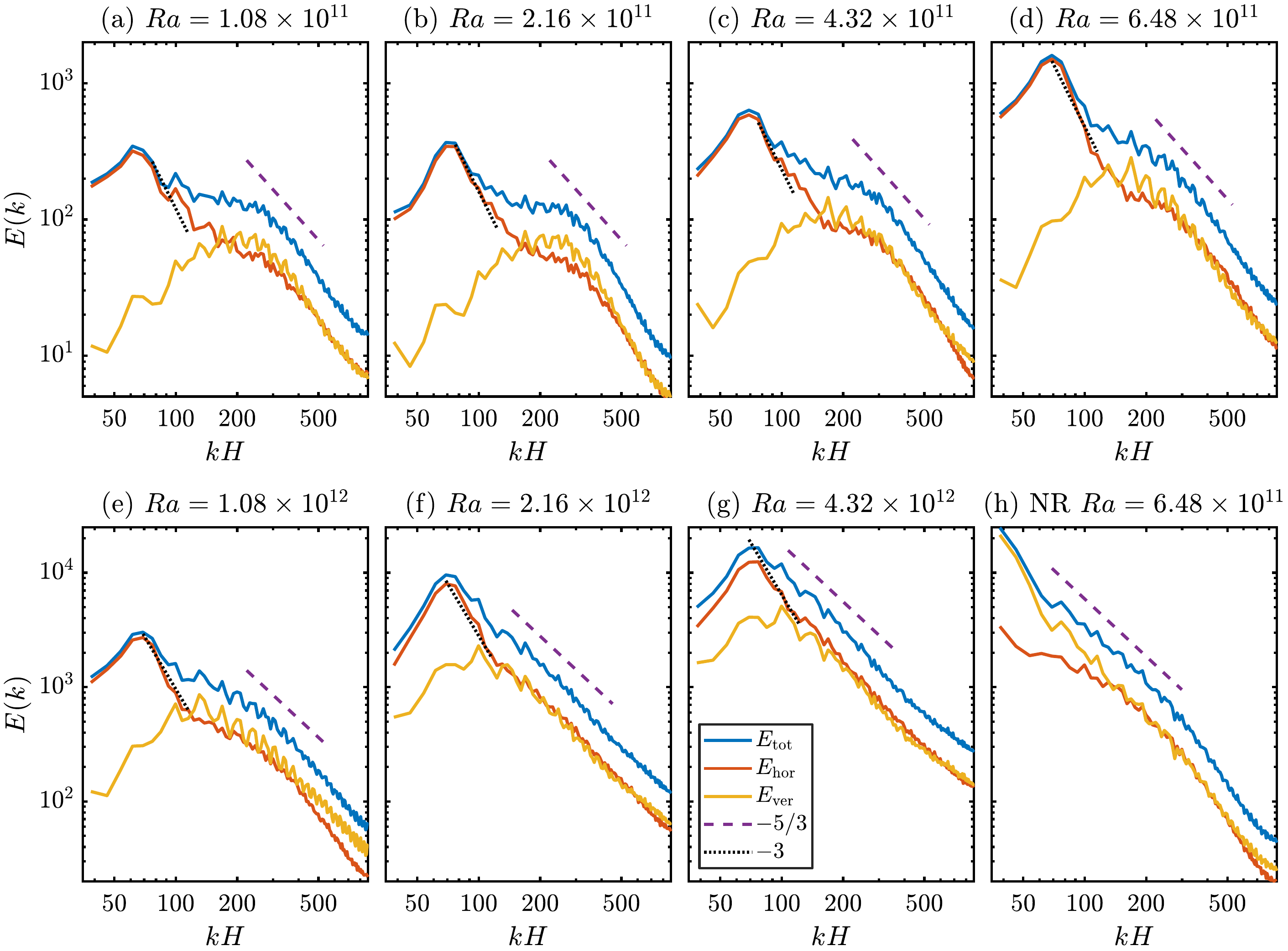}
\caption{\label{fi:spectra}Kinetic energy spectra $E(k)$ plotted as a function of normalised wavenumber $kH$, including total kinetic enery ($E_\mathrm{tot}$) as well as the contributions from horizontal ($E_\mathrm{hor}$) and vertical ($E_\mathrm{ver}$) velocity components. Reference power-law slopes $k^{-5/3}$ and $k^{-3}$ are also included.}
\end{figure}

In figure \ref{fi:spectra} we show the energy spectra of our seven RRBC cases as well as the non-rotating case. We plot the total energy spectrum ($E_\mathrm{tot}$) together with the respective contributions from the horizontal ($E_\mathrm{hor}$) and  vertical ($E_\mathrm{ver}$) velocity components. Wavenumbers are normalised using the cell height $H$.

The rotating cases show common features. In all of them, $E_\mathrm{tot}$ exhibits a peak at $kH \approx 70$, which corresponds to the physical wavelength $L \approx 0.18 \unit{m}\approx D/2$. This peak is related to the ordering into a quadrupolar vortex state that we reported before \citep{mack21}, which in hindsight was also found in the mean flow fields of \citet{wamcck20}. \citet{ezs22} also show this flow organisation in their figure 2(c). Jets pointing radially inward are observed at the positions where up- and downward flowing parts of the sidewall circulation meet, resulting in a four-quadrant organisation with two cyclonic and two anticyclonic vortices. The formation of these jets is a nonlinear feature of the sidewall circulation absent from its linear description \citep{hb93,lzc06,zl09}. The dependence on the geometrical aspect ratio $\Gamma$ is an open question; we do anticipate changes for wider cylinders where more than one azimuthal wavelength can fit into the circumference. Here, with one azimuthal wavelength, we see an organisation into a quadrupolar vortex where each pole has a characteristic size $L\approx D/2$. These structures are most energetic, but at the same time they are quasi-two-dimensional: there is no signature of them in the spectra $E_\mathrm{ver}$ based on vertical velocity.

The shape of the remaining part of each spectrum is strongly dependent on the vertical energy contribution. As $Ra$ increases, the peak in $E_\mathrm{ver}$ goes towards smaller $k$ (larger length scales), a feature that we have observed before \citep{mack21} from calculations of the characteristic horizontal correlation length scale based on vertical velocity. We interpret this as the scale at which buoyancy is most prominently adding energy into the turbulent motion. From the maximum of $E_\mathrm{ver}$ towards larger $k$ values, we see the gradual development of a spectral scaling range close to $E(k)\sim k^{-5/3}$, the predicted scaling for the inertial range of three-dimensional turbulence \citep{f95,p00}. This scaling range is most prominent in the $E_\mathrm{tot}$ curves and at larger $Ra$. The sum of horizontal and vertical contributions is clearly reflected in the total spectra $E_\mathrm{tot}$, with the peak due to the quadrupolar vortex dominating at smaller $k$ and a broad shoulder due to the peak contribution of $E_\mathrm{ver}$ that shifts as a function of $Ra$.

The horizontal spectra, in fact, also show a different scaling trend, from the approximate location of the peak of $E_\mathrm{ver}$ to smaller wavenumbers ending on the maximum of $E_\mathrm{hor}$, that resembles $E(k)\sim k^{-3}$ scaling. This $k^{-3}$ scaling could be a sign of a (nonlocal) inverse cascade \citep{sw99,l07,jrgk12,amcock20}, although this scaling range is short and certainly cannot be considered as a proof. We are investigating the energy transfer among scales in more detail.

The nonrotating RBC case, instead, shows a very different situation. The peak of the overall spectrum is at $k_\mathrm{min}$, and it is dominantly due to the contribution of $E_\mathrm{ver}$. It is a sign of the presence of the large-scale circulation of nonrotating convection \citep[e.g.,][]{agl09}: a domain-filling vertical convection roll with one half of the domain displaying upward flow and the other half downward. We have observed it in the current measurements \citep{mack21,m22}. This structure corresponds to a wavelength of approximately the diameter of the cylinder, that in wavenumber space is translated to $k_\mathrm{min}$. Here we also observe a rather extensive scaling range nicely matching the theoretical prediction $k^{-5/3}$.

\section{\label{ch:concl}Conclusion}
We present flow statistics that result from the stereo-PIV measurements of rotating Rayleigh--B\'enard convection (RRBC) in the geostrophic regime. We analyse radial profiles of the root-mean-square (rms) values of horizontal and vertical velocity as well as vertical vorticity. Near the sidewall, the radial extent of the sidewall circulation can be clearly distinguished. We see that for both velocities and vorticities higher $Ra$ corresponds to larger values of these quantities, and at the same $Ra$ the non-rotating case exhibits larger magnitudes than the corresponding rotating one.

Within the bulk region, for the flow states from plumes to rotation-influenced turbulence (RIT), the Reynolds number scalings follow a trend that lies in the range of both CIA and VAC scaling predictions, perhaps being numerically closer to the former, but otherwise no decisive distinction can be made. The vorticity scaling follows the same trend as the horizontal velocity, indicating that the characteristic horizontal scale remains constant \citep[e.g.,][]{sjkw06,jrgk12,mack21}. We also find that energy is distributed nearly isotropically, in contrast to nonrotating RBC where the vertical velocity component is strongly preferred.

The Reynolds number results compared to previously published results reveal that there is an overall trend towards ``ultimate" diffusion-free scaling \citep{ahj20} as the Ekman number is reduced, but we do not reach it yet. However, there are indications of convergence in scaling for results coming from different methods (experiments, direct numerical simulations and asymptotically reduced models) and different geometries (confined cylinders and horizontally periodic computational domains). At lower $Ra$ there are complications due to the use of a confined geometry, as it promotes the presence of a wall mode that affects the bulk flow statistics.

Spectra of kinetic energy provide further evidence for the formation of a prominent quadrupolar vortex structure \citep{mack21}. This feature is remarkably two-dimensional, as it is reflected only in spectra of horizontal velocity. In the spectra of vertical velocity we observe a broad peak, that we associate with the dominant length scale for energy injection. With increasing $Ra$, the peak shifts towards smaller wavenumbers \citep[larger length scales;][]{mack21}. For larger wavenumbers we observe the development of a scaling range with a classical $-5/3$ slope for three-dimensional turbulence \citep{f95,p00}. This inertial-range scaling is most prominent for the nonrotating case.

These results provide valuable input on the turbulence scaling of flows in the geostrophic regime, a property that is scarcely investigated up to now. There is clear perspective in the observed convergence towards diffusion-free scaling. At the same time, to reach true convergence even more extreme parameter values than those used here are likely required, which is a major challenge for both experiments and direct numerical simulations. Additionally, the influence of the strong convection mode near the sidewall, the sidewall circulation, should be further explored to ensure proper interpretation of experimental and numerical results from confined domains.

%\backsection[Supplementary data]{\label{SupMat}Supplementary material and movies are available at \\https://doi.org/10.1017/jfm.2019...}

%\backsection[Acknowledgements]{Acknowledgements may be included at the end of the paper, before the References section or any appendices. Several anonymous individuals are thanked for contributions to these instructions.}

\backsection[Funding]{M.M., A.J.A.G. and R.P.J.K. received funding from the European Research Council (ERC) under the European Union's Horizon 2020 research and innovation programme (Grant Agreement No. 678634). We are grateful for the support of the Netherlands Organisation for Scientific Research (NWO) for the use of supercomputer facilities (Cartesius and Snellius) under Grants No. 2019.005, No. 2020.009 and No. 2021.009.}

\backsection[Declaration of interests]{The authors report no conflict of interest.}

\backsection[Data availability statement]{The data that support the findings of this study are available from the corresponding author, R.P.J.K., upon reasonable request.}

\backsection[Author ORCIDs]{Matteo Madonia, https://orcid.org/0000-0002-8510-8103; Andr\'es J. Aguirre Guzm\'an, https://orcid.org/0000-0002-4942-5216; Herman J.H. Clercx, https://orcid.org/0000-0001-8769-0435; Rudie P.J. Kunnen, https://orcid.org/0000-0002-1219-694X.}

%\backsection[Author contributions]{Authors may include details of the contributions made by each author to the manuscript'}

\end{document}